\documentclass[a4paper,12pt]{article}
\usepackage[latin1]{inputenc}
\usepackage{graphicx, amssymb, amsfonts, amsmath, mathrsfs}
\usepackage[english]{babel}
\usepackage[official]{eurosym}
\usepackage{caption}[2008/04/01]
\usepackage{newcent}
\usepackage{rotating}
\usepackage{lineno}
\usepackage{setspace}
\usepackage{tabularx}
\usepackage[flushmargin]{footmisc} %%%%%%pas d'indentation des footnotes%%%
\usepackage{color}
\usepackage{lscape,array,hhline,pstricks,fancyhdr}
\usepackage{lineno}
\usepackage[hscale=0.76,vscale=0.80]{geometry}
\usepackage{geometry}
\usepackage{booktabs}
\usepackage{array}
\usepackage{chngpage}
\usepackage{url}

\newcommand{\PreserveBackslash}[1]{\let\temp=\\#1\let\\=\temp}

\newcolumntype{C}[1]{>{\PreserveBackslash\centering}p{#1}}

\newcolumntype{R}[1]{>{\PreserveBackslash\raggedleft}p{#1}}

\newcolumntype{L}[1]{>{\PreserveBackslash\raggedright}p{#1}}

\title{\LARGE{\textbf{ China building energy consumption: definitions and measures from an operational perspective}}}

\author{Ling-Yun HE $^{1,2,3}$ and  Wei WEI$^{4,}$ \footnote{ denotes the corresponding author. Dr. HE is a full professor of energy economics and environmental policies. WEI is a research associate supervised by Dr. HE.  The authors contribute equally in the project. HE conceived the whole project. WEI calculated and analysed the results under Dr. HE's supervision. HE and WEI co-wrote the manuscript. The authors would like to thank all our colleagues from both China Agricultural University and JiNan University, for all their warm helps, constructive suggestions and pertinent comments. This project is supported by the National Natural Science Foundation of China (Grant Nos. 71273261 and 71573258), and China National Social Science Foundation (No. 15ZDA054).}       \\\small 1. Institute of Resource, Environment and Sustainable Development Research, \\ \small JiNan University, Guangzhou 510632, China \\ \small 2. Department of International Economics and Trade, School of Economics, \\ \small JiNan University, Guangzhou 510632, China\\ \small 3. College of Economics and Management, Nanjing University of Information Science and Technology,\\ \small Nanjing 210044, China\\ \small 4. College of Economics and Management, China Agricultural University,\\ \small Beijing 100083, China\\
 \small * Corresponding author.
\\ \small Emails: lyhe@amss.ac.cn}

\date{\small Submitted on \today}

\begin{document}
\maketitle

\begin{abstract}
\noindent
There is an increasing awareness of the significance of Chinese building energy consumption (BEC). However, something worth discussing is that estimate the building energy consumption adopting the definition of life cycle or operation. In the existing studies with various evaluation methods, the issue about the amount of energy consumed by China buildings has not been understood. In order to settle the disputes over the calculation of BEC, this paper establish an appropriate accounting method of building energy to present BEC situation in China and lay the foundation for building energy efficiency. Adopting the conception of building operational energy consumption, we find that the energy consumption of buildings just accounts for $15\% - 16\%$ of the final total energy consumption in China; by contrast, the previous calculations  usually have double accounting through top-down approach if central heat-supply of buildings was given into additional consideration.
\end{abstract}

\\
\\

\begin{onehalfspace}
\section{Introduction}\label{section_intro}

\begin{figure}[h]

\includegraphics[width=16.5cm]{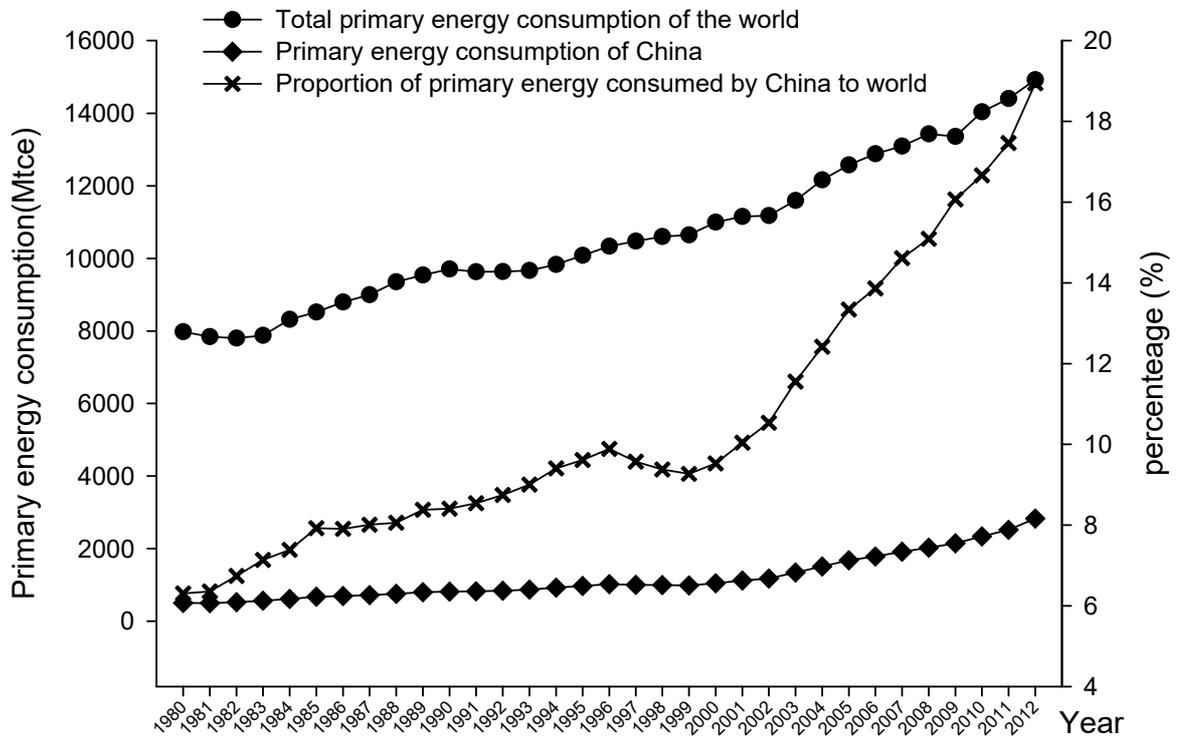}\\
\caption{The total primary energy consumption during 1980 to 2012}
\label{Fig.1}
\end{figure}

As one of the biggest countries of energy consumption, China catches global eyes and its energy use condition has been regarded as a crucial problem of the energy security. The \emph{International Energy Outlook 2016} (IEO2016) \footnote{Source: \url{http://www.eia.gov/forecasts/ieo/index.cfm}}shows rising levels of energy demand over the next three decades, especially led by strong increases in Asian countries outside of the Organization for Economic Cooperation and Development (OECD). Moreover, the outlook predicts that non-OECD Asia, including China and India, account for more than half of the world¡¯s total increase in energy consumption over the 2012 to 2040 projection period. For the past several decades, There has been consistent growth in total primary energy consumption throughout the 226 countries and regions. At the same time, the primary energy consumption of China is not only increasing, but its proportion of the total primary energy consumption has tripled, jumped to 18.94 per cent from 6.30 per cent(see Fig.\ref{Fig.1})\footnote{Source: \url{ http://www.eia.gov/cfapps/ipdbproject/IEDIndex3.cfm?tid=44&pid=44&aid=2}}. Thus more and more attentions as well as concerns are taken to energy consumption situation in China, which is a matter of world energy demand.

Lombard et al.(2008) hold the idea that the energy consumption from building fall between 20\% and 40\% in developed countries, which far exceeds the ratios of the energy consumed in the industrial and transportation sectors. Building energy consumption in China, the largest developing country in the world, can be featured by super high absolute demand. It is equivalent to the total energy consumption in the Middle East, is twice the consumption level in Africa, and even the sum of consumption of Japan and South Korea(Lin, 2015). It follows that building energy consumption of China has impact not only domestically but also internationally.

\begin{figure}[h]
\includegraphics[width=16.5cm]{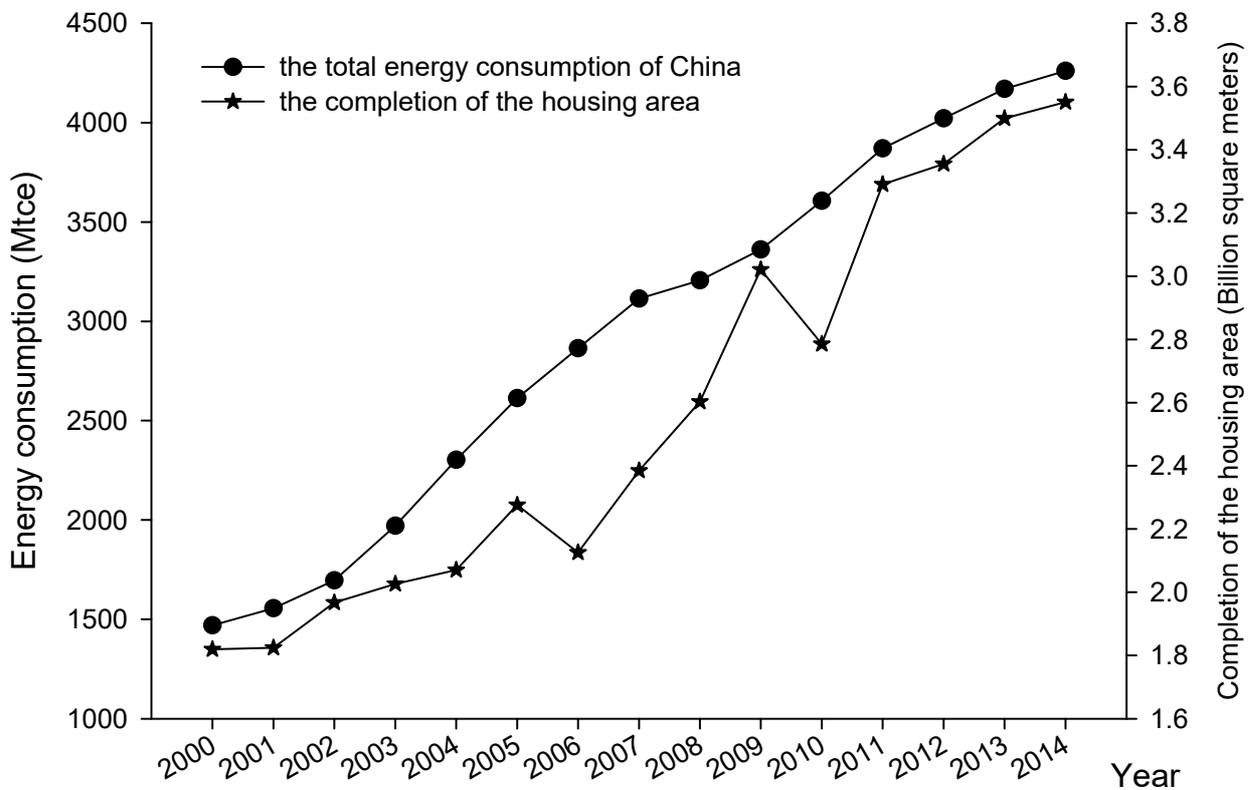}\\
\caption{The Energy Consumption and the building completion area in China during 2000 to 2014}
\label{Fig.2}
\end{figure}

According to the National Bureau of Statistics (NBS) of China, from the year of 2000 to 2014, the total energy consumption of China has been scaled up to 4160 million metric tons of coal(Mtce), and keeps increasing dramatically (see Fig.\ref{Fig.2}). As the rapid urbanization, together with the boom of construction industry, billions of square meters of housing were completed year by year from 2000 to 2014 (Fig.\ref{Fig.2}). Besides, the living standard of the people improved lead to a increase in the amount of  household appliances. In a word, the percentage of building energy consumption with respect to the total national energy consumption may rise higher for the urbanization and the request to living quality.

For guaranteeing energy security and tackling climate problem caused by energy consumption, building energy is an issue worthy of attention. From September 22nd to 25th, in 2015, Chinese president Xi, during his state visit to the United States, emphasized on promoting the construction of the China-U.S. Energy Efficiency Fund project, which is aimed at investment in China's building energy efficiency projects\footnote{Chinese, U.S. companies launch fund to promote energy efficient buildings. News, Global Edition. (2015). Available from: \url{http://news.xinhuanet.com/english/2015-09/17/c_134633961.htm} }. Although building energy consumption has caused the widespread concern of government and nation, the actualities of building energy consumption still remain to be clarified.

Through analyzing the building energy consumption situation in China, one of the most critical problem to be solved is that the amount of energy has been exactly consumed by China building. There is sharp differences in the statistics of building energy consumption among previous researches. It was once a relatively broad acceptance that the building energy consumption accounted for 27.4\% of total energy consumption, but Yang, Wei and Jiang(2007) find out that in that calculation energy consumption of construction industry is regarded as building energy consumption, and there still exist vague  between the concept of intermediate and final energy consumption. As the table.\ref{Table.1} show, there's a wide variety in the ratio of building energy consumption depending on different definition, approach and calculation items.
\begin{table}[hp]
  \caption{The building energy consumption situation of different statistics in China}
  \label{Table.1}
 \begin{tabular}{L{3cm}L{3cm}L{5cm}L{2cm}L{2cm}}
  \toprule
  Article  &Definition& Calculation items &Approach& Proportion \\
  \midrule
  (Long, 2005)&energy consumed by maintenance function and operation of buildings& heating and other energy use &top-down & about 20\% \\
  (Building Energy Research Center of Tsinghua University, 2013) & civil building energy use of operating period & heating in North Urban China, commercial and public buildings(except central-heating), urban residential buildings(except central-heating), rural residential buildings & bottom-up& 20.6\%(in 2011)  \\
  (Wang, 2007)  & civil building energy use of operating period  & energy used in civil, commercial and other departments of Energy Balance of China (except petroleum products consumed by transportation) &top-down&19.7\%(in 2004), 20.7\%(in 2005)\\
  (Zhang, Lu, and Ni, 2008)   & civil building energy use of operating period& residential buildings and public buildings &bottom-up&13.77\%(in 2004)\\
  (Li and Jiang, 2006)& energy used by the life cycle of building & energy consumption of building operation, energy consumption of building materials production and indirect energy consumption of buildings &top-down& average 45.5\% (during 1998 to 2003)\\
  (Zhang, He, Tang, and Wei, 2015)& energy used by the life cycle of building& building materials production energy, construction energy, operation energy, energy savings from recycling &top-down & about 46\% (during 2001 to 2013)\\
  \bottomrule
  \end{tabular}
\end{table}

By comparison, the gap of amount of building energy firstly comes from the distinct definitions. Secondly, the modeling methods can fall into two categories, the top-down and the down-top approaches, in existed studies on estimating building energy consumption. Then the disunity accounting calibers, that items are calculated in appropriate model, also lead to the diversity of results. In final, due to the limit of existed statistical system, it is difficult to pick up a certain amount of energy use that should belong to building, or to deduct a certain amount of energy that consumed by other industries. Thus a more immediate problem is that establishing a appropriate method definitely reveals the actualities of building energy consumption as much as possible.

The deviation in calculation and definition of building energy consumption would directly influence the understanding of the energy situation in China. So a statistical method, that applying to top-down approach primarily explores the interaction between the energy sector and macro economy, need to be established. Meanwhile, the method can eliminate mixed consumption from other sectors, so that its results can provide sufficient data and countermeasures for building energy efficiency work in China. This paper adopts the definition of building operation energy to clearly evaluate the benefits of policies. The building life-cycle assessment mixes energy use of construction, materials, and operation together so that it is hard to make targeted measures for energy saving.

Given the importance of the building energy and the divergent results on the amount of building energy consumption in the existed literature, the findings of the current study are excepted to enrich this pool of unifying accounting method surrounding building energy consumption and to give the energy consumption of buildings close to the fact. This study proposes that there exists double counting among previous studies, and that the extra energy consumption is combined into building energy consumption owing to the life-cycle perspective. The objective of this study is to develop a statistical method of energy only consumed by building running. In particular, even if the method underestimates the amount of building operating energy, the part of the ambiguity is eliminated.

To achieve these objectives, the model tailored to calculate China's building energy consumption on a macroscopic view is established, from which a clear idea about current situation of energy consumption related to buildings in China is got. The following section (Section 2) compares the two definitions of building energy consumption and section 3 establishes a proper model estimating building operational energy consumption. The rest of the paper is arranged as follows. The energy consumption situations of buildings in China during 2000 - 2013 is revealed by employed the given model in section 4, around the results above, characteristics of building energy consumption of China and the reasonableness of the established model are further discussed and analysis in Section 5. Finally, the central conclusions are shown in the last part, Section 6.

\section{The analysis of concept of building energy consumption}

\subsection{The definition of building energy consumption}
There are two kinds of definition of building energy consumption in the existed researches. The one is the operating energy which is expended in maintaining indoor environment and performing building's equipment such as heating and cooling, air conditioning,  lighting, household appliances, office equipment, hot water supply, cooking, elevator, ventilation, and other operating appliances. The another one is life cycle energy of buildings. The total life cycle energy of buildings include both embodied energy and operating energy . The embodied energy refers to be sequestered in building materials during all processes of production, on-site construction, and final demolition and disposal (Dixit, 2010). It means that all energy is used in buildings' life cycle, such as energy used in the production of construction materials, buildings construction, buildings operational phase and buildings demolition.
\subsection{The distinction between operating energy and life cycle energy}
The operational energy is consumed by all activities related to the usage of the buildings. The operating energy consumption is influenced by the using habits of the occupants and the efficiency of equipment in the buildings. while the embodied energy is expended in the initial construction, maintenance, renovation and demolition. The embodied energy consumption depends on reducing material use, selection of materials with a lower embodied carbon and energy intensity (Ibn-Mohammed, 2013). The embodied energy is buried in the entire process of buildings' utilization. The embodied energy can be saved by improving production and technology, reducing energy intensity and so on.

At present, the embodied energy has caused more and more concerns and the definition of life cycle energy is applied to the analysis of building energy consumption. However, operating energy and embodied energy are all covered by the life cycle energy, so that the life-cycle building energy consumption conceals the energy-saving effect of  efficient equipment and appliances. Meanwhile, the policy efficiency of energy conservation aiming at operating stage is obscured by embodied energy. In addition, the current emphasis shifting to embodied energy attracts attention on industrial technology improvement, it ignores the potential for curbing operating energy. So it is difficult to distinguish the policy efficiency of either operational energy conservation or embodied energy conservation by the statistical method of life-cycle building energy consumption. In order to a clear and integrated image of building energy consumption for  policymakers, building operational energy consumption is adopt to establishes the statistical method in this study.
\section{The statistical method of the building operational energy consumption}

\subsection{The type of building }
In China, the buildings can be classified as the industrial buildings and the civil buildings, and the civil buildings are divided into the residential buildings and the public buildings. The public buildings include the office buildings, commercial buildings (shopping malls, emporiums), tourism buildings (hotels, entertainment), buildings of science and education, cultural and health, communication buildings (telecommunications, communications, broadcasting station) and transportation buildings (airport, station construction)(table.\ref{Table.2}).

\begin{table}[htb]
  \caption{The classification and the definitions of civil buildings in China}
  \label{Table.2}
 \begin{tabular}{L{5.5cm}L{11cm}}
  \toprule
  Types of buildings  & Definitions \\
  \midrule
  The civil building        & General term of the buildings for people living and public buildings \\
  The residential building  & The buildings for people living  \\
  The public building       & The buildings for people of various public activities \\

  \bottomrule
  \end{tabular}
\end{table}

According to the definition of operating energy consumption and the classification of national economic industries, we can find that four industries in the Energy Balance Sheet of China are involved in the energy consumption of the civil buildings. They are Transport, Storage and Post, Wholesale, Retail Trade and Hotel, Restaurants, Residential Consumption (Urban, Rural) and Others (table.\ref{Table.3}).
\begin{table}[htb]
  \caption{The industry of energy balance sheet involved in building energy consumption}
  \label{Table.3}
 \begin{tabular}{L{5.5cm}L{5.5cm}l}
 \toprule
  Type of building&        &Industry   \\
 \midrule
     Public buildings & Commercial buildings    & Wholesale, Retail Trade   \\
                      & Transportation buildings& Transport, Storage and Post \\
                      &Tourism buildings         & Hotel, Restaurants \\
                      &Office buildings and others & Others \\
     Residential buildings &                          & Residential Consumption \\
\bottomrule
  \end{tabular}
\end{table}

\subsection{The calculation model }
Statistical reporting system is implemented for Chinese energy consumption collecting. Firstly, by the Statistics Law of the People's Republic of China, State Statistical Bureau develops the system of energy statistics. The system of energy statistics include the production, purchase, consumption and inventory of main energy products. Secondly, various provinces and cities determine for themselves the investigation method of Energy Balance Sheet and Consumption of Energy by Sector. Finally, State Statistical Bureau collects and gathers the all data submitted by provinces, autonomous regions and municipalities.

When the data of energy consumption was collected, because of the system of energy statistics, the energy consumption which includes the building energy use is all submitted by business enterprises and units. The building energy consumption is mixed together into the energy consumption of various industries. If using the existing macro-statistics, the building energy consumption need to be decoupled from industries of Energy Balance Sheet.

Across all industries of the Energy Balance Sheet, it is disputed that whether calculate the energy consumption of industrial buildings as building energy. On one hand, from the perspective of building operation, comparing to the energy use of industrial production, the amount of industrial building operating energy is small to find. On the other hand, accounting the building energy use is aimed to lay the foundation for the pursuit of effective energy-saving potentials. The industrial buildings have the main purpose that is
the production, the enormous amount of energy consumption in the industrial buildings is out of the control of building energy efficiency.
the energy consumption of industrial buildings should not include the energy use engaged in industrial production. Besides, from the view of definition, it is accepted that the building energy consumption is the energy consumption of the civil building in the process of use by international practice(Wang, 2007). Therefore the energy use of the industrial buildings is not on our list in this paper.

The civil buildings are categorized into the residential buildings and the public buildings. The energy, used in the residential buildings, is consumed by temperature adjustment, lighting, cooking, household appliances inside the buildings, etc. Rural residential building and urban residential building have different ways of energy consumption. Moreover, the most of living fuel in the residential building of China rural area usually depends on non-commercial energy resources such as straw, firewood, etc. The energy use of public buildings is scattered in various sectors. Besides, concentration heat-supply of the  buildings is usually taken into consideration in previous studies for its vast energy consumption. However, in this paper, the energy consumption of central heating is not extra added to the energy consumption of buildings, because if adopting the Energy Balance of China, the central heating has been included. Then non-commercial energy resources in rural areas are also considered. With the combination of the system of energy statistics and the existing data resources, the model of residential buildings energy consumption can be expressed as
\begin{equation}
NBE=RE+PE+NCE
\end{equation}
where NBE refers to national total energy consumption in the building operation process; RE represents energy used in the residential buildings; PE refers to energy consumption of the public buildings; NCE indicates the non-commercial energy resources consumed in the rural areas.

\subsection{Data}
The model of building energy consumption in China established above is determined by means of the statistical method and available data. The data applied in this article are stem from China Statistical Yearbooks and Energy Balance of China Energy Statistical Yearbooks. China Statistical Yearbooks can be found on the web site of National Bureau of Statistics of China \footnote{Source: \url{http://www.stats.gov.cn/}}. China Energy Statistical Yearbooks are accessible on the web site of China economic and social development statistics database\footnote{Source: \url{http://tongji.cnki.net/kns55/Navi/NaviDefault.aspx}}.

\section{Results}

\subsection{Residential building energy }
The residential buildings can  provide a place for people's living, involving villa, dormitory, apartment, etc. The residential consumption denotes that in the daily life, consumption is to meet the needs of residents own and their family members. So the consumption in the residential buildings, such as cooking, heating, is also the part of residential consumption. In terms of Energy Balance Sheet, the residential buildings energy is mainly distributed in residential consumption sector. Considering the accessibility of data information and previous studies, energy consumption of residential buildings can be calculated by the following formula:

\begin{equation}
\label{equation2}
RE=RC-GRC-95\%\times DRC
\end{equation}

where RE represents energy consumed in the residential buildings; RC refers to final energy of residential consumption, that is from China's Energy Balance Table; GRC denotes that gasoline is used in the residential consumption; DRC indicates the diesel oil consumed in residential daily life. Wang(2007) stated that the oil of private automobiles was accounted for the main consumption of petroleum products in residents' life, so all of gasoline and ninety-five percent of diesel oil are used by transportation in the residential consumption. In this paper, access to net building energy consumption, we completely remove the oil product for use in transportation by the following equation:
\begin{equation}
\label{equation3}
RE=RC-GRC-DRC
\end{equation}

\begin{figure}[htbp]
\includegraphics[width=16.5cm]{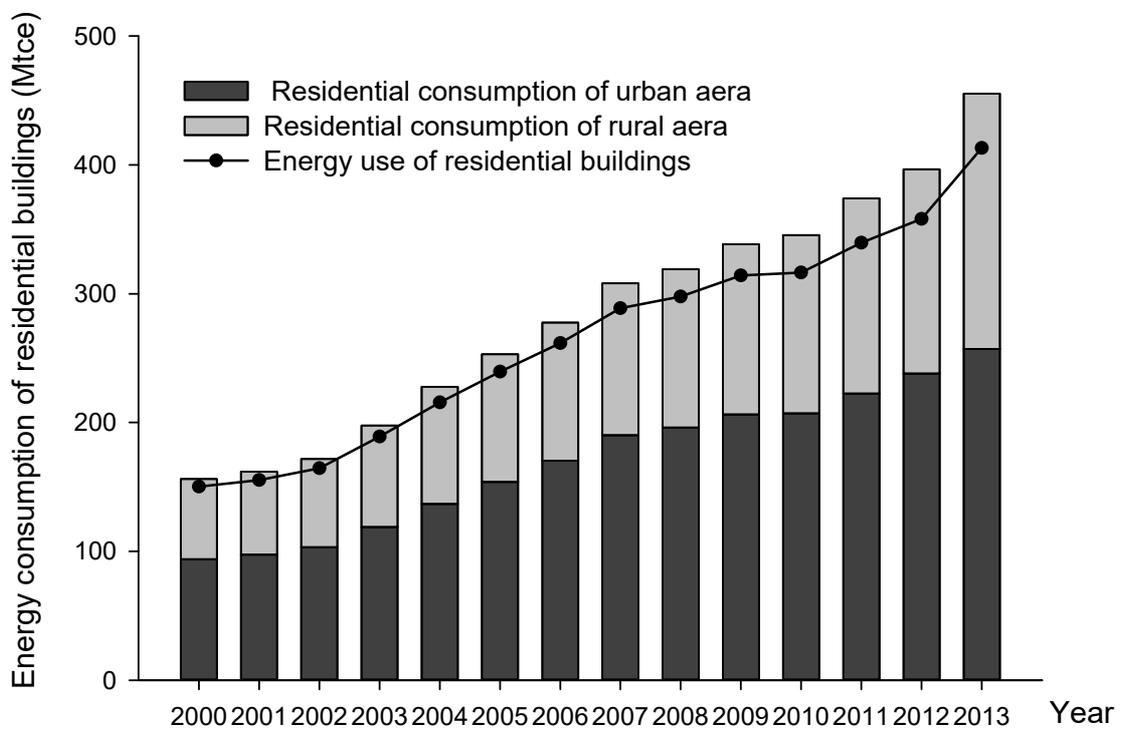}\\
\caption{The energy use of residential buildings in China during 2000 to 2013}
\label{Fig.3}
\end{figure}

On the grounds of formula Eq.\ref{equation3}, Fig.\ref{Fig.3} presents the energy consumption of residential buildings with traffic energy fully deducted for the period of 2000 - 2013. Obviously, the upward trend of energy consumption in residential buildings is continuing during 2000 to 2013, that is partly in association with the rapid urbanization and improvement of living quality. In addition, the energy consumed in the residential buildings of rural area in China increases rapidly, that is because commercial energy resource is popularized in the countryside.
\subsection{Public building energy}
In light of the fact that the public buildings exist in all fields, by National Economical Industry Classification, we collect the energy,used in the public buildings, from Transport, Storage and Post, Wholesale, Retail Trade and Hotel, Restaurants, and others of energy balance sheet.
\subsubsection{Transport, Storage and Post industry }
In the traditional method of accounting, electric power consumption of Transport, Storage and Post industry (Table.\ref{Table.4}) was deemed to make contribution to the energy use of public buildings. However, With the increasing application of high-speed EMU (Electric Multiple Units), the electricity consumption of EMU contributes the most part in electric power consumption of Transport, Storage and Post industry. For the case of lack of data, we calculate the energy, consumed by public building of Transport, Storage and Post industry, by including the electricity consumption or not.
\begin{table}[htbp]
  \caption{The electric power consumption of Transport, Storage and Post industry in China $(2000-2013)$ }
  \label{Table.4}
 \begin{tabular}{cc}
 \toprule
    Year&  \  \ Electricity consumption of Transport, Storage and Post industry (Mtce)  \\
 \midrule
     2000&3.46 \\
     2001&3.80\\
     2002&3.72\\
     2003&5.00\\
     2004&5.53\\
     2005&5.29\\
     2006&5.74\\
     2007&6.54\\
     2008&7.03\\
     2009&7.58\\
     2010&9.03\\
     2011&10.43\\
     2012&11.25\\
     2013&12.30\\
\bottomrule
  \end{tabular}
\end{table}
 \subsubsection{Wholesale, Retail Trade, Hotel, Restaurants, and others industry}
Similarly, it is necessary to exclude both gasoline and diesel oil from the energy use of Wholesale, Retail Trade, Hotel, Restaurants, and others industries. Then, the remaining energy consumption is considered as the operation energy of the public buildings in the above several industries. The previous studies (Wang, 2007) have shown that 95\% of gasoline, 35\% of diesel oil in services industry is used in transportation. But the estimation was still subject to be verified. For guarantee the purity of buildings energy, both gasoline and diesel oil are all excluded from the industrial final energy consumption. The results are shown in Table.\ref{Table.5}.
\begin{table}[htbp]
\footnotesize
\caption{Energy consumption of Wholesale, Retail Trade, Hotel, Restaurants, and others industries in China $(2000-2013)$ }
  \label{Table.5}

 \begin{tabular}{p{.05\textwidth}p{.2\textwidth}p{.2\textwidth}p{.12\textwidth}p{.12\textwidth}p{.25\textwidth}}
 \toprule
    Year& Energy consumption of WRHR (Mtce) &  Energy consumption of others industry (Mtce) & Gasoline consumption (Mtce) &  Diesel oil consumption (Mtce) & Buildings consumption of WRHR and others (Mtce) \\

 \midrule
     2000&30.48&57.62&12.69&10.70&64.70\\
     2001&31.70&59.32&12.85&11.25&66.92\\
     2002&33.73&62.41&13.84&12.40&69.91\\
     2003&39.15&71.53&14.05&13.05&83.58\\
     2004&44.84&82.43&16.28&14.96&96.02\\
     2005&48.48&92.55&16.59&14.73&109.70\\
     2006&53.14&102.76&17.47&15.49&122.94\\
     2007&56.89&111.58&18.41&16.63&133.43\\
     2008&57.34&117.71&18.50&19.01&137.54\\
     2009&64.12&126.90&17.91&19.14&153.97\\
     2010&68.27&136.81&19.63&21.62&163.82\\
     2011&77.95&151.89&21.93&23.90&184.01\\
     2012&85.46&165.81&24.43&24.39&202.45\\
     2013&105.98&197.63&30.01&22.92&250.67\\
\bottomrule
  \end{tabular}
\scriptsize
 note: WRHR represents Wholesale, Retail Trade, Hotel, Restaurants industry.
\end{table}
\subsubsection{Total public buildings energy}
\begin{figure}[htbp]
\includegraphics[width=16.5cm]{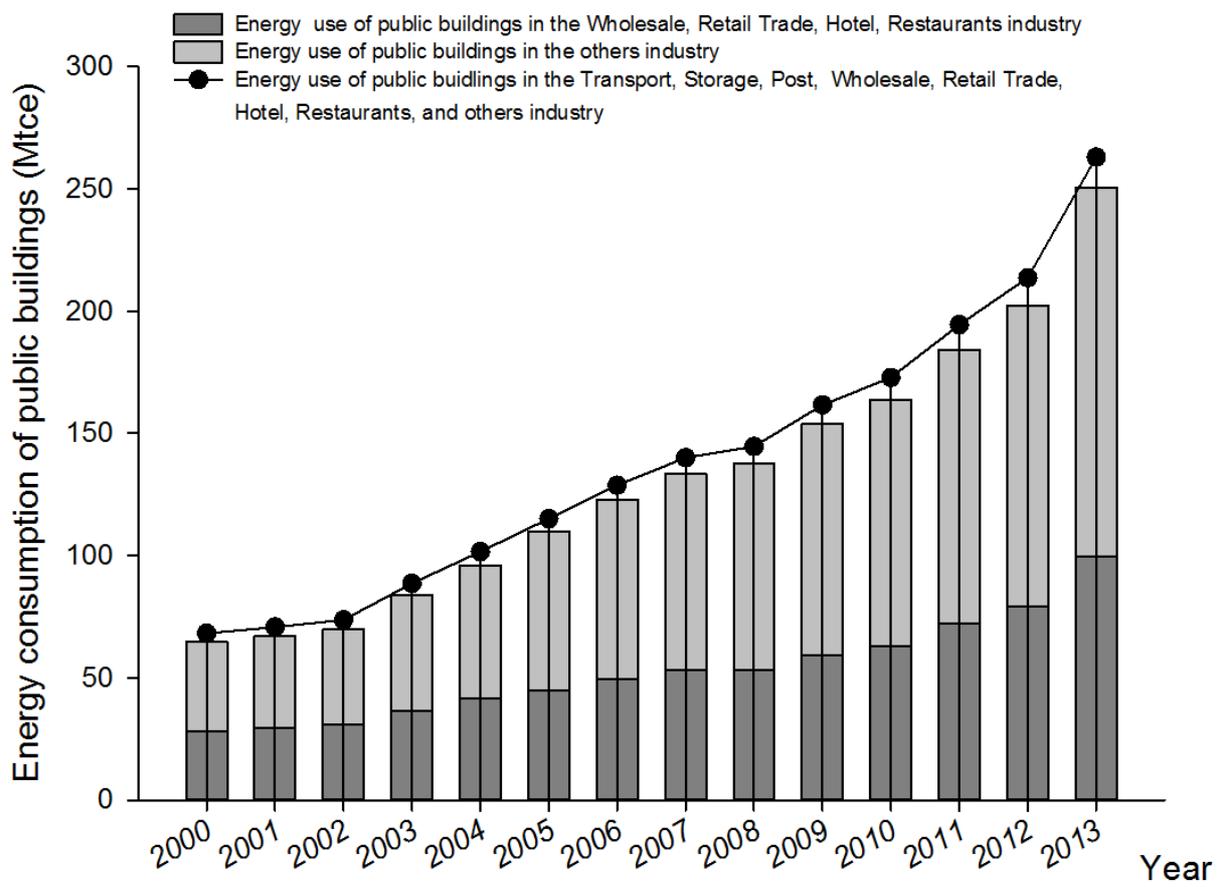}\\
\caption{The energy consumption of public buildings in China during 2000 to 2013}
\label{Fig.4}
\end{figure}

The total energy consumption of public buildings in China can be obtained by adding up the values of energy consumption in preceding two parts. If only calculating energy consumption of Wholesale, Retail Trade, Hotel, Restaurants and others industry, the public building energy would be underestimated. For purposes of comparison, the different results are shown in Fig.\ref{Fig.4} .
\subsection{Non-commercial energy}
In general, commercial energy refers that national energy products whose the whole or most enters into the commodity market for transactions, such as fossil fuels(coal, oil and natural gas), nuclear energy, and hydropower. While the energy products, self-produced, self-collected, self-consumed, is non-commercial energy such as fuel wood, straws and methane. Because Energy Balance of China does not include non-commercial energy, this paper, exclusively or additionally, collect the non-commercial energy consumption in the rural area of China. The components proportion of non-commercial energy consumption in rural buildings is considerable in view of Table.\ref{Table.6}.
\begin{table}[htbp]
\small
\caption{Non-commercial energy consumption of rural area in China $(2000-2013)$ }
  \label{Table.6}

 \begin{tabular}{p{.1\textwidth}p{.3\textwidth}p{.28\textwidth}p{.3\textwidth}}
 \toprule
    Year& Consumption of fuel wood and straw (Mtce) &  Consumption of methane (Mtce) & Total non-commercial energy consumption (Mtce) \\
\midrule
     2000&204.12&1.62&205.74\\
     2001&228.38&2.20&230.58\\
     2002&255.49&2.68&258.17\\
     2003&259.19&3.30&262.49\\
     2004&266.23&3.99&270.22\\
     2005&262.69&4.93&267.62\\
     2006&274.76&5.09&279.85\\
     2007&252.69&7.31&260.01\\
     2008&221.29&8.45&229.74\\
     2009&189.88&9.34&199.22\\
     2010&158.48&9.97&168.44\\
     2011&127.07&10.91&137.98\\
     2012&95.66&11.84&107.50\\
     2013&64.26&12.77&77.03\\
\bottomrule
  \end{tabular}
\end{table}
\subsection{Total uilding energy}
\begin{figure}[t]
\includegraphics[width=16.5cm]{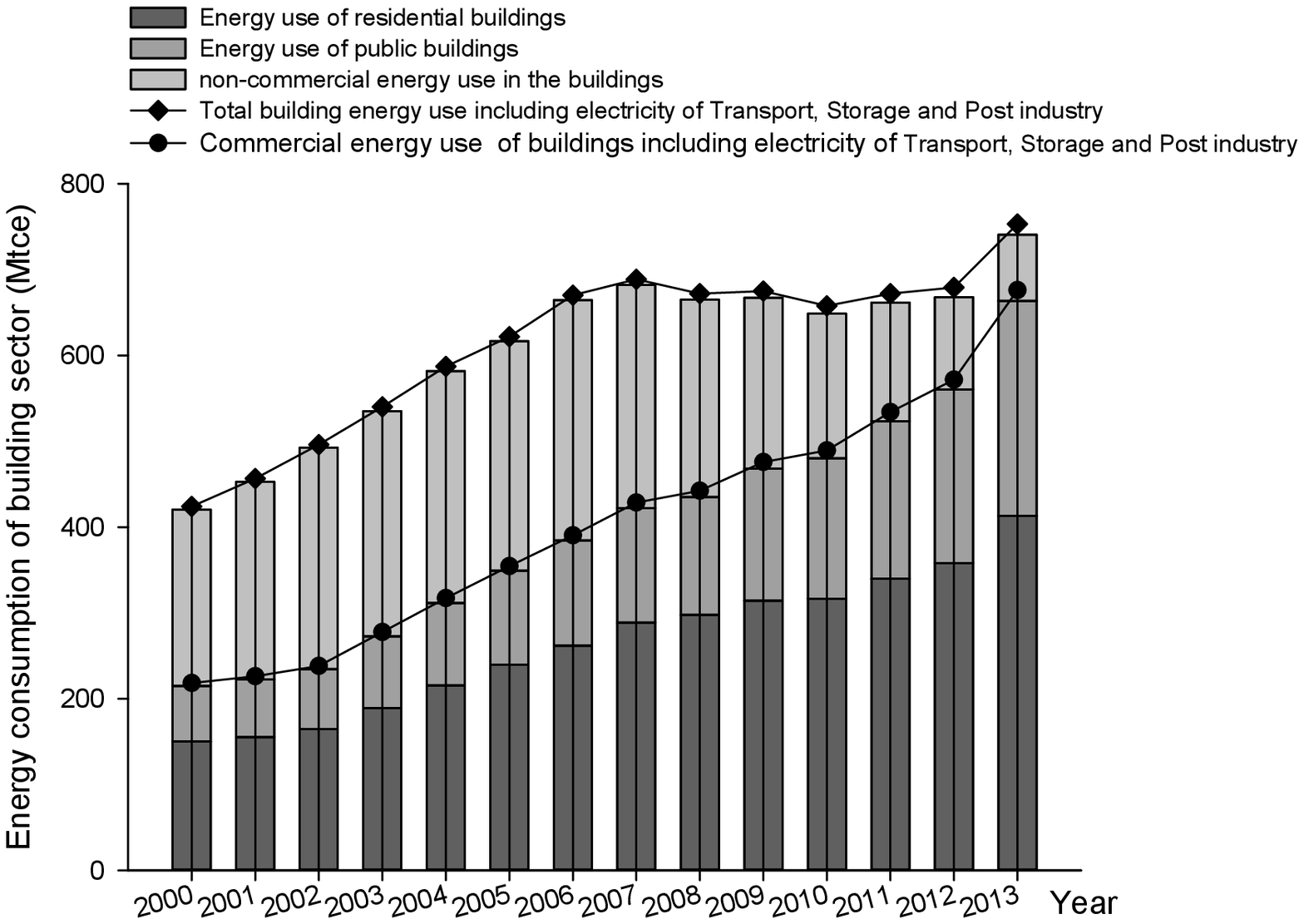}\\
\caption{The total energy consumption in China's building sector $(2000-2013)$}
\label{Fig.5}
\end{figure}

The building energy consumption, consumed in the running process, can be determined by aggregating the results of 4.1$\-$4.3. China's building energy consumption, refers to commercial energy, has experienced an uninterrupted growth from 218.34 Mtce in 2000 to 676.06 Mtce in 2013, as shown in Fig.\ref{Fig.5}
\section{Discussion and analysis}
\subsection{National building energy consumption}
\begin{figure}[t]
\includegraphics[width=16.5cm]{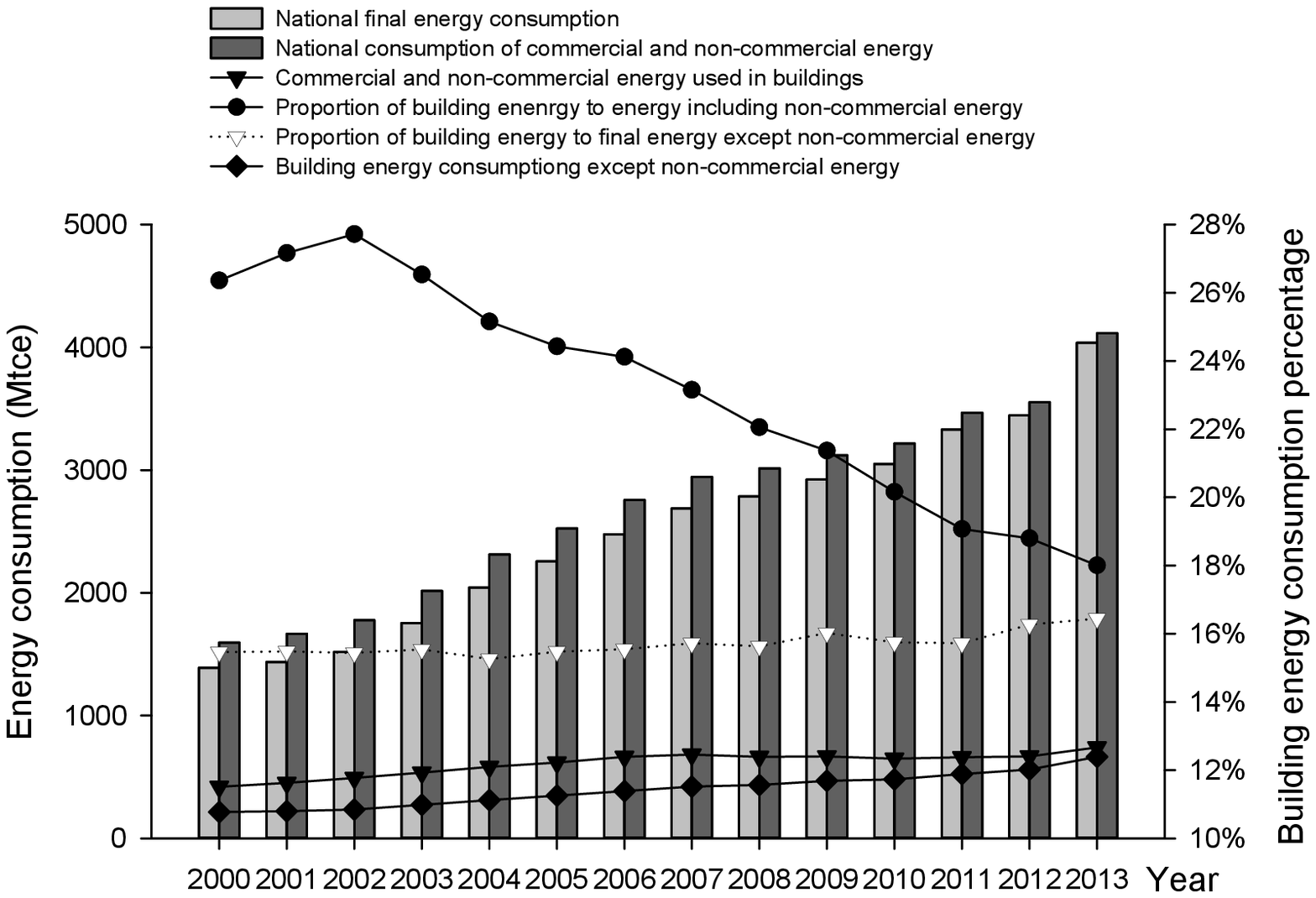}\\
\caption{Energy consumption of buildings and its proportion in China $(2000-2013)$}
\label{Fig.6}
\end{figure}

The electricity consumption of Transport, Storage and Post industry is excluded to the national building energy, because it is hard to allocate the part of buildings' use from the electricity consumption on the basis of existing data. Fig.\ref{Fig.6} presents the energy consumption of buildings in comparison with China's final energy consumption and final energy consumption including non-commercial energy over the period of 2000 - 2013.

The final energy consumption refers to the last part of energy consumption or energy obtained by the entrance of terminal energy equipment. According to China's statistics system, the energy consumption of various industries is collected from the final energy used by all sectors in Energy Balance of China. However, the energy statistics of Energy Balance Sheet just list commercial energy. The final energy including non-commercial energy of rural area is shown in Fig.\ref{Fig.6}. Obviously, the non-commercial energy consumption is decreasing with popularization of commercial energy and cost improvement of access to non-commercial energy, so that the proportion of building energy energy to final energy including non-commercial experiences a declining trend. To explore the interaction between energy consumption and economy, the top-down approach is applied. The definition of building operating energy is benefit to investigate energy-saving behaviors and habits. In fact, the net building consumption accounts for merely 15-16 per cent with respect to the final energy consumption in China. Overall, the total amount of building energy makes foundation of building energy conservation. It is significantly important to pay more attention to know the building sector¡¯s energy situation and , based on that, to achieve the target of energy saving and emission reduction and ensure energy security of China.

\subsection{Energy consumption of central heating in buildings }
In terms of energy composition, coal, petroleum products, natural gas, liquefied natural gas, heat, electricity and other energy would be captured in the Energy Balance of China. As the heat has already been included in the final energy, it is unnecessary that the energy consumption associated with central-heating  is counted in the building energy consumption. Take China's Energy Balance Sheet (2013) for example shown in Table.\ref{Table.7}. From the table, the energy supplied by heat-supply enterprise has been added in the final energy, moreover we can find out that the most heating supply is used by Industry. Therefore, applying top-down approach, the energy consumption of central heating calculated again results in excessive amount of building energy which doesn't accord with the truth.
\begin{table}[h]
\small
\caption{Part of Energy Balance of China (Standard Quantity) - 2013 }
  \label{Table.7}

 \begin{tabular}{p{.8\textwidth}p{.2\textwidth}}
 \toprule
    Item & Heat (Mtce) \\
\midrule
     \textbf {Input (-) or Output (+) of Transformation } & \textbf{124.33}  \\
     \ \ Thermal Power&-16.54\\
     \ \ Heating Supply&123.48 \\
     \ \ Recovery of Energy&17.39 \\
     \textbf {Loss}& \textbf{1.43}\\
     \textbf {Total Final Consumption}& \textbf{122.90}\\
     \ \ Agriculture, Forestry, Animal Husbandry, Fishery and Water Conservancy&0.04\\
     \ \ Industry&89.21\\
     \ \ Construction&0.27\\
     \ \ Transport, Storage and Post&0.78\\
     \ \ Wholesale, Retail Trade and Hotel, Restaurants&1.73\\
     \ \ Others&3.09\\
     \ \ Residential Consumption&27.78\\
\bottomrule
  \end{tabular}
\end{table}
\subsection{Structural composition of energy consumption in buildings}
\begin{figure}[h]
\includegraphics[width=16.5cm]{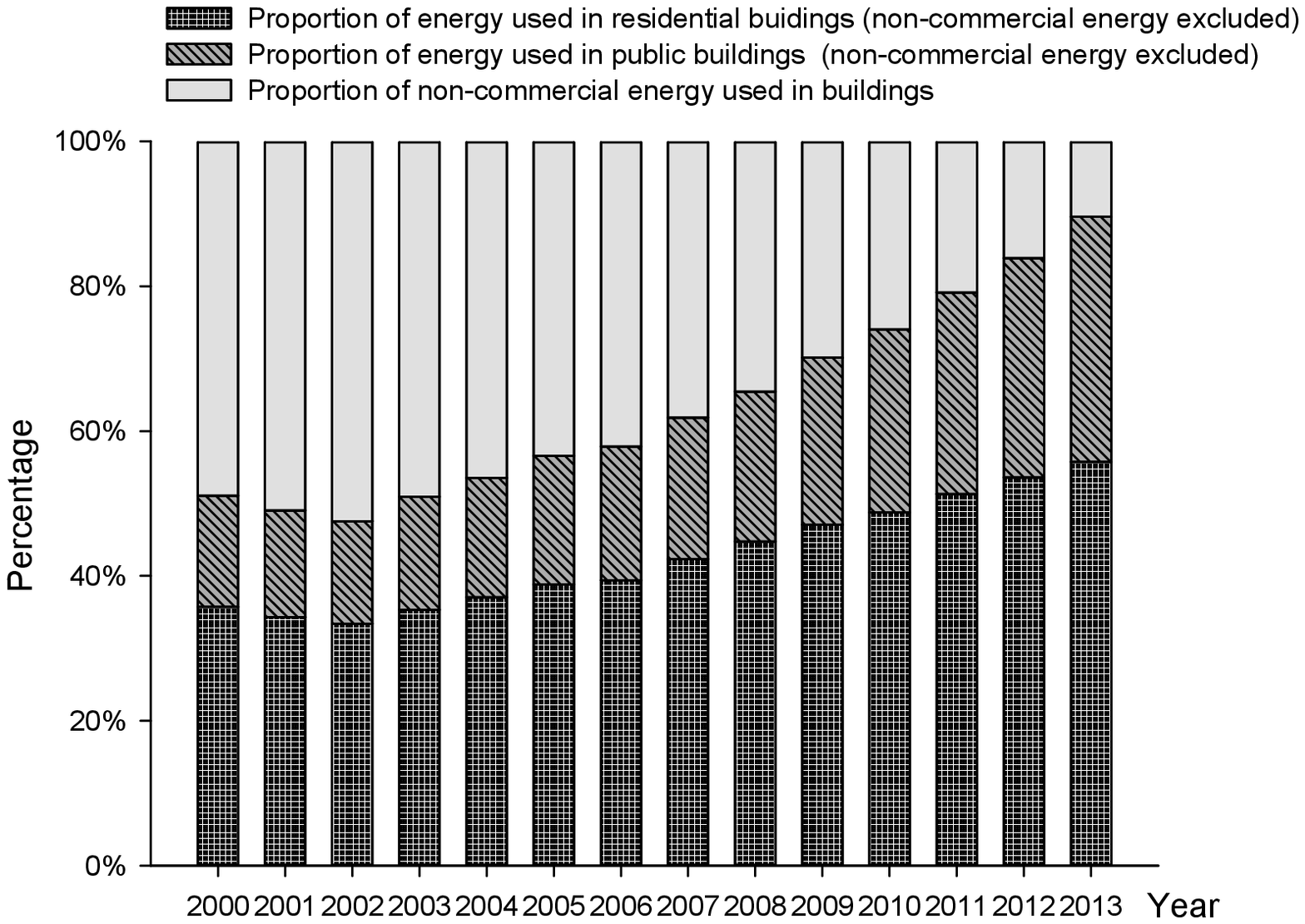}\\
\caption{Energy consumption of buildings and its proportion in China $(2000-2013)$}
\label{Fig.7}
\end{figure}

Fig.\ref{Fig.7} depicts the composition of China's energy consumption in the buildings from 2000 to 2013. Obviously, residential buildings energy and non-commercial energy predominate in building energy. The two proportions present exactly opposite variation trends. As more commercial energy has been used in rural areas, the percentage of non-commercial energy decreases continuously, having dropped to 10.40\% in 2013. Specifically, the proportion of public building energy experiences a upward trend, gradually increased from 15.38\% in 2000 to 33.84\% in 2013, it is a potential field to effectively reduce energy consumption in future.

The pursuit for better life quality and living condition in recent years leads to the rapid expansion of building energy use, especially the residential building energy consumption. With regard to the energy usage habit, how to keep balance between energy saving and welfare of the residents remains further study.

\subsection{The environmental effects of building energy}
Buildings are major consumers of energy. Only the energy consumption of residential building sector ranked second after the industry sector (Chen, 2008). The environmental effects of energy consumption from 15 per cent or 45 per cent of the final energy lie between two extremes. By improving building energy efficiency, the European Community research has found that carbon emissions from buildings would be reduced by 22 per cent \footnote{EU directive on the energy performance of buildings. UK: Department for Environment, Food $\&$ Rural Affairs.(2008). Available from: \url{http://www.defra.gov.uk/ ENVIRONMENT/energy/internat/ecbuildings.htm.}}. Promoting building energy efficiency will help achieving goals for reducing carbon emissions as well as improve the energy performance of new and existing buildings. The overvalued building energy will mislead the judgement of its Carbon and even pollution emission. So that the environmental problems associated with building energy consumption is widely misunderstood.

\section{Conclusion}
\paragraph{}The huge energy consumption of China challenges its energy security and sustainable development along with increasing desire for indoor comfort. Meanwhile in order to perform the duties of reducing emission, building energy efficiency is expected to play a significant role in addressing energy problems in China. However, the blurry data about buildings energy consumption brings blindness to the national building energy conservation policy. Therefore, data verification of building energy consumption becomes extremely urgent. In this paper, it is more clear that building energy can be calculated by the definition of operating energy. what's more, it's easier to assess the environmental effects of the net building energy consumption. Without redundance, the building energy is to provide useful information with policymakers.

A model of building energy is established in this paper to response the controversy and disputes to the exact amount of building energy consumption. To ensure energy security and sustainable development, building energy consumption is expected to become a target of energy conservation. Therefore this study has been presented to determine the net energy consumption in buildings from building operating perspective. By the comprehensive analysis carried out above. we obtain the following conclusions:\begin{enumerate}
                               \item China¡¯s building energy consumption without redundance only accounts for 15-16 per cent of the final energy, rather than about 45 per cent, that is caused by including the energy use of material production, construction and so on. The life cycle building energy is not benefit to the policy targeted at conservation in a stage of  building life cycle.
                               \item Consideration of the policy relevance, the blurry data is eliminated as much as possible. Specially, the central heat-supply of buildings that was repeated accounted in prior studies is deducted.
                              \item  The residential building energy consumption dominate the use of building energy in China, the major conservation is improving the habit of energy usage. The percentage of public building energy consumption to building energy is also increasing. while the non-commercial energy consumption reduce obviously.
                             \end{enumerate}

Above all, the building energy consumption is underestimated, but under the condition that redundance and imprecise part are deducted, all that remained is consumed by building. Meanwhile, more efficient amount of building energy consumption is collected and shared, with regard to lay the foundation for energy-saving in China.
\section*{References}

\noindent

{\footnotesize
Building Energy Research Center, Tsinghua University. (2013)2013 Annual Report on China Building Energy Efficiency, China Building Industry Press, Beijing, China.

Cai, W. G., Wu, Y., Zhong, Y., $\&$ Ren, H. (2009). China building energy consumption: situation, challenges and corresponding measures. \emph{Energy policy}, 37(6), 2054-2059.

Chen, S., Li, N., Guan, J., Xie, Y., Sun, F., $\&$ Ni, J. (2008). A statistical method to investigate national energy consumption in the residential building sector of china. \emph{Energy and Buildings}, 40(4), 654-665.

Dixit, M. K., Fern\'{a}ndez-Sol\'{i}s, J. L., Lavy, S., $\&$ Culp, C. H. (2010). Identification of parameters for embodied energy measurement: a literature review. \emph{Energy and Buildings}, 42(8), 1238-1247.

Geng Q, NBS, (2010). \emph{China Manuals on Energy Statistics}.

Li, Z.J., $\&$ Jiang, Y., (2006). Pondering over the Situation of Domestic Generalized Building Energy Consumption. \emph{Architectural Journal}(7), 30-33(in Chinese).

Lin, B., $\&$ Liu, H. (2015). China's building energy efficiency and urbanization. \emph{Energy and Buildings}, 86, 356-365.

Long, W. D., (2005). Building energy consumption ratio and building energy efficiency target. \emph{China Energy}, 27(10), 23-27(in Chinese).

Ibn-Mohammed, T., Greenough, R., Taylor, S., Ozawa-Meida, L., $\&$ Acquaye, A. (2013). Operational vs. embodied emissions in buildings - a review of current trends. \emph{Energy and Buildings}, 66(5), 232¨C245.

Ministry of Construction of the PRC, (2005), \emph{Code for design of civil buildings}.

NBS, China Statistical Database, 2000-2013 \url{http://data.stats.gov.cn/workspace/index?m=hgnd}

NBS, China Energy Statistics Yearbook, 2000-2013 \url{http://tongji.cnki.net/kns55/Navi/HomePage.aspx?id=N2010080088$\&$name=YCXME$\&$floor=1}

Peng, C., Yan, D., Guo, S., Hu, S., $\&$ Jiang, Y. (2015). Building energy use in China: Ceiling and scenario. \emph{Energy and Buildings}, 102, 307-316.

P\'{e}rez-Lombard, L., Ortiz, J., $\&$ Pout, C. (2008). A review on buildings energy consumption information. \emph{Energy and buildings}, 40(3), 394-398.

Ramesh, T., Prakash, R., $\&$ Shukla, K. K. (2010). Life cycle energy analysis of buildings: an overview. \emph{Energy and Buildings}, 42(10), 1592-1600.

Wang, Q.Y.(2007). Research on statistics and calculation of building energy consumption in China. \emph{Energy Conservation and Environmental Protection}, (8): 9-10 (in Chinese).

Yang X, Wei Q.P, Jiang Y. (2007). Study on statistical method for building energy consumption, \emph{Building Energy Efficiency} 35 (1), 7¨C10 (in Chinese).

Zhang, H.B., Lu, S.H., Ni, D.L., (2008). Research on statistical model and method of building energy consumption. \emph{Building Science}, 24(8), 19-24 (in Chinese).

Zhang, Y., He, C. Q., Tang, B. J., $\&$ Wei, Y. M. (2015). China's energy consumption in the building sector: A life cycle approach. \emph{Energy and Buildings}, 94, 240-251.

}

\end{onehalfspace}
\end{document}